\documentstyle[prbbib,aps]{revtex}
\begin{document}
{\Large
\draft
\title{
"SPHERICAL"~ 3--STATE POTTS SPIN GLASS: EXACT SOLUTION
}
\author{
N. V. Gribova, V.N.Ryzhov, and E.E.Tareyeva
}
\address{
Institute for High Pressure Physics, Russian Academy of Sciences, Troitsk
142190, Moscow region, Russia}
\date{\today}
\maketitle
\maketitle
\begin{abstract}
A continuous 3-state Potts model with an analog of spherical
constraints is proposed and is shown to have an exact solution
in the case of infinite-ranged interactions. "Spherical"~ 3-state
Potts spin glass model is solved using the known properties of a
large random matrix. For this model the results are identical to
those obtained by the replica approach for replica symmetric
solution.
\end{abstract}

During last decades the so called 1RSB (one--step
replica--symmetry breaking) models: the $p$--state Potts spin
glasses (PG) and $p$--spin spin glasses as well as their
continuous versions -- are in the
focus of investigations in spin glass theory. Although the
spherical models are rather unrealistic they present rare
examples of many--particle systems which can be solved
analitically in three dimensions. It is well known that the
behaviour of continuous magnetic systems with spherical
constraints differs essentially of their discrete counterparts
being in some aspects nonphysical. Nevertheless it seems always
interesting to consider models which have exact solutions in
regular as well as in random couplings cases.

In this Letter we introduce and solve a new model -- "spherical"~
continuous generalization of three--state Potts model. This can
be easily done due to a representation of 3-state Potts model in
terms of quadrupole moment operators used in our papers dealing
with the 3-state Potts spin glass. The content of this Letter is
the straightforward generalization of the paper by Berlin and
Kac~\cite{BerK} where the spherical two-spin
model was introduced and solved for nearest neighbour
interactions on different lattices and that by Kosterlitz,
Thouless and Jones ~\cite{KTJ} where this model was solved in
the case of infinite-ranged interactions with a Gaussian
probability distribution (spherical version of the
Sherrington-Kirkpatrick (SK)~\cite{SK} model).

The $p$-state Potts spin glass model is a lattice model where
each lattice site carries a Potts spin $\sigma_i$ which can take
one of the $p$ values $\sigma_i = 0, 1, ..., p-1$ with the
Hamiltonian
\begin{equation}
H=- \frac{p}{2} \sum_{i \not= j}  J_{ij} \delta_{\sigma_i
\sigma_j}
\label{ham}
\end{equation}
where $\delta_{\alpha \beta}$ is the Kronecker symbol. Thus, a
pair $\{\sigma_i,\sigma _j\}$ contributes an energy $-J_{ij}$
if $\sigma _i = \sigma _j$ and zero otherwise. The interactions
$J_{ij}$ are quenched random variables described by a Gaussian
distribution
$$P(J_{ij}) = ( \sqrt {2 \pi} J)^{-1} \exp
{[-(J_{ij}-J_0)^{2}/2J^2]}.$$
The Potts glass with an infinite--range interaction
$J_0= \tilde J_0/N$, $J= \tilde J/N^{1/2}$ has been studied in
~\cite{EldSh83,LaEr83,Eld84,LaNu84,gks,LT1,LT2,LT3,LT4,sapari94}.
The short--range version has been considered in
~\cite{Gold85,cwi1,cwi2} and is a subject of intense
investigation through computer simulations ~\cite{Binder}.

The soft--spin version  of PG has also been suggested
as a starting point for a theory of structural glasses and the
transition from the metastable fluid to the glass state
~\cite{kirkp}. The Potts glass may also serve as a model for
orientational glasses in molecular crystals and cluster glasses
where a strong single--site anisotropy restricts the orientation
of the appropriate molecular group to $p$ distinct directions.

As is well known in SK discrete spin glass ~\cite{SK}
the replica symmetric (RS) solution is unstable and one
needs to use the Parisi scheme to obtain the
stable solution with so called full replica symmetry
breaking (FRSB). In the corresponding spherical spin glass
the exact solution can be obtained by straightforward thermal
averaging and using the known properties of a large random
matrix and this solution is identical to RS solution obtained
through replica approach ~\cite{KTJ}.

In the case of $p$-spin spherical spin glass with $p>2$
it was
established rigorously by
Crisanti and Sommers ~\cite{som}
that there is
a discontinuous transition to 1RSB solution
and this solution remains stable till zero
temperature.
For discrete Ising spin $p$-spin
glass E.Gardner ~\cite{gard} has shown that 1RSB solution is
unstable at very low temperature using perturbations about
known solutions for the cases $p=2+\epsilon $ and $p\to\infty$.
The second transition leads to a phase described by a continuous
order parameter  full RSB (FRSB) function $q(x)$.

The 3-state Potts spin glass is somehow intermediate system
between SK ($p$=2) glass and "canonical"~ 1RSB glasses.
In 3-state PG there is no reflection symmetry and 1RSB solution
was shown ~\cite{gks,LT3} to be stable in the vicinity of the
RS transition temperature (which coincides with that of
1RSB transition) against the higher stages of RSB, but the static
and dynamical transitions are both continuous, which is not a
general property of 1RSB models. Recently, it was shown
~\cite{GRT} that at low temperature the 1RSB solution becomes
unstable and a 2RSB or full RSB can take place.

Let us consider now the system of particles on lattice sites
$i,j$ with the reduced Hamiltonian ~\cite{LT1,LT2}:
\begin{equation}
H=- \frac{1}{2} \sum_{i \not= j}  J_{ij} (Q_i Q_j + V_i V_j),
\label{ham1}
\end{equation}
where $Q=3 J^2_{z} - 2$, {\bf J}=1, $J_z=1, 0,-1$.
A particle quadrupole moment is the second-rank tensorial
 operator with five components. In the principal axes frame only
two of them remain: $Q$ and $V$. In the subspace {\bf J}=1 the
following equality holds:
$$ \frac {1}{6} (Q_{m(i)} Q_{n(j)} + V_{m(i)} V_{n(j)} + 2) = \delta _{mn}.$$
This equality shows the equivalence of ~(\ref{ham1}) to the
$p=3$ Potts Hamiltonian ~(\ref{ham}).
We shall assume that
$J_{ij}$ are
distributed following the
Gaussian law with zero mean:  $$P(J_{ij}) = ( \sqrt {2 \pi}
J)^{-1} \exp {[-J_{ij}^{2}/2J^2]},$$ and $J= \tilde J/N^{1/2}$.

It is easy to show that the operators $Q$ and $V$ in the
subspace $J=1$ are such that $QV = V$, $Q^2 = 2-Q$ and $V^2 = 2
+ Q$ so that
\begin{equation} Q^2 + V^2 = 4.
\label{qv}
\end{equation}
This equality is the key one permitting to introduce a
continuous "spherical"~ generalization of the discrete 3-state
Potts model. The Hamiltonian~(\ref{ham1}) with the constraints
\begin{equation}
\sum_{i}(Q_i^2 + V_i^2) = 4N
\label{sph}
\end{equation}
can be solved exactly and the solution occurs to be identical to
the replica symmetric solution obtained through replica
approach.

Let us first diagonalize the matrix $J_{ij}$ by an orthogonal
transformation. The variables $Q_i$ and $V_i$ are transformed to
new variables $Q_{\lambda}$ and $V_{\lambda}$ defined by
\begin{equation} Q_{\lambda} = \sum_{i} <\lambda|i>Q_i;~~~
V_{\lambda} = \sum_{i} <\lambda|i>V_i,
\label{slam}
\end{equation}
where $<\lambda|i>$ is the orthonormal eigenvector of $J_{ij}$
belonging to the eigenvalue $J_{\lambda }$. In the limit
$N\to\infty$ the eigenvalue density $\rho (J_{\lambda })$
obeys the semicircular law (see e.g.~\cite{meh}):
\begin{equation}
\rho (J_{\lambda }) = \frac{1}{2 \pi \tilde J ^2} (4 \tilde J ^2
- J_{\lambda }^2)^{1/2}.
\label{meh}
\end{equation}
The constraints ~(\ref{sph}) on the integration in
partition function may be taken into account by use of
corresponding $\delta $-function. Using integral representation
for $\delta $-function and performing integration over the
variables $Q_{\lambda }$ and $V_{\lambda }$ as is usually done
in spherical-model approach we obtain the partition sum in the
form (up to unimportant normalization factor):
\begin{equation}
Z = \int\limits_{a - i \infty}^{a + i \infty} \frac
{dz} {2 \pi i} \exp{[N(4z - \frac {1}{N} \sum_{\lambda }
\ln(z - 2 \beta J_{\lambda }))}],
\label{zz}
\end{equation}
where the contour of integration is to the right of the largest
eigenvalue, $2 \tilde J$.

Using the integration $ \int dJ \rho (J)...$ with $\rho (J)$ from
~(\ref{meh}) instead of
the summation $(1/N) \sum_{\lambda }...$ we obtain the saddle-point
equation for $z$:
\begin{equation}
z - \sqrt {z^2 - t^2} = 2 t^2,
\label{sad}
\end{equation}
where $t = \beta \tilde J$. The equation~(\ref{sad}) has the
solution $z= \frac{1}{4}(1 + 4 t^2)$ if $T>T_c$ and $T_c$ is
determined by the condition $z^2 = t^2$, that is $4 t_c^2 = 1$.
For $T<T_c$, Eq.~(\ref{sad}) has no solution, and the
saddle-point value of $z$ sticks at $t$, the branch point of the
integrand of Eq.~(\ref{zz}) (see ~\cite{KTJ}).
The partition function ~(\ref{zz}) gives the free energy per
site averaged over the eigenvalue spectrum ~(\cite{meh})
$(k_B=1)$:
\begin{equation}
<f(T)>_{av} = -T -2 \frac{\tilde J^2}{T} -2 T \ln {2}, T>T_c,
\label{fav1}
\end{equation}

\begin{equation}
<f(T)>_{av} = -4 \tilde J + \frac{T}{2} + T \ln {\frac{\tilde
J}{2T}}, T<T_c.
\label{fav2}
\end{equation}

Specific heat per site is $4 t^2$ for $T>T_c$
and $1$ for $T<T_c$. The low--temperature entropy is negative
and diverges logarithmically as $T\to 0$. This nonphysical
behaviour is analogous to the standard pathology of spherical
models.

Let us consider now the Hamiltonian ~(\ref{ham1}) using replica
approach. Using standard procedure we can write the replica free
energy per site in the following form (compare with ~\cite{LT1,LT2,GRT}:
$$
-\beta f= \lim_{N \to \infty} \lim_{n \to 0}
\frac{1}{nN}
\biggl\{
\int\limits_{a -i\infty}^{a +i\infty}
dz
\left[
\int dQ_{1}...dQ_{n}
dV_{1}...dV_{n} dy_1 dy_2 dx_1 dx_2 \times \right.$$
$$\left. \times
\exp{\left\{-\frac{n(n-1)y_1^2}{4}
-\frac{n(n-1)y_2^2}{4}-\frac{nx_1^2}{2}
-\frac{nx_2^2}{2}+4nz-z\sum_{\alpha }(Q_{\alpha }^2
+V_{\alpha }^2)\right\}} \times \right.$$\\
$$\left. \times \exp {\left\{
\frac{ty_1}{2}\sum_{\alpha \neq \beta }Q_{\alpha
}Q_{\beta } +
\frac{ty_2}{2}\sum_{\alpha \neq \beta }V_{\alpha
}V_{\beta } +
\frac{t x_1}{\sqrt {2}}\sum_{\alpha }Q_{\alpha }^2+
\frac{t x_2}{\sqrt {2}}\sum_{\alpha }V_{\alpha }^2 \right\}}\right]^N -
1\biggr\}$$
Here $x_1 \sim <Q_{\alpha }^2>$, $x_2 \sim <V_{\alpha
}^2>$, $y_1 \sim <Q_{\alpha} Q_{\beta }>$,
$y_2 \sim <V_{\alpha} V_{\beta }>$.

Taking into account that

$$\int dQ_{\alpha _1}...dQ_{\alpha _n} \exp {\left\{
\frac{ty_1}{2}\sum_{\alpha \neq \beta }Q_{\alpha
}Q_{\beta } +
\frac{t x_1}{\sqrt {2}}\sum_{\alpha }Q_{\alpha }^2 -z\sum_{\alpha
}(Q_{\alpha }^2\right\}} = $$
$$=\exp {\left[ -\frac{n}{2} \ln
{(z-\frac{tx_1}{\sqrt {2}}+\frac{ty_1}{2})} + n \frac {ty_1}{4
(z-\frac{tx_1}{\sqrt {2}}+\frac{ty_1}{2}}\right])}$$
we obtain finally the saddle-point free energy in the form:
\begin{equation}
- \beta f=  4z -\frac{x_1^2}{2} - \frac{x_2^2}{2} +
\frac{y_1^2}{4} + \frac{y_2^2}{4} - \frac{1}{2} \ln
{(z- \frac{x_1 t}{\sqrt {2}} + \frac {y_1 t}{2})} +
 \label{frep}
\end{equation}
$$+\frac{ty_1}{4( z-\frac{x_1 t}{\sqrt {2}} + \frac {y_1 t}{2})} -
\frac{1}{2}\ln {(z- \frac{x_2 t}{\sqrt {2}} + \frac {y_2 t}{2})}
+ \frac{t y_2} {4(z- \frac{x_2 t}{\sqrt {2}} + \frac {y_2 t}{2})}$$

It is easy to see that the equations for the extremum condition
for ~(\ref{frep}) have a solution $x_1 = x_2 = x$ and $y_1 = y_2
= y$. Now these equations have the form:
\begin{equation} 8(z-
\frac{x t}{\sqrt {2}} + \frac {y t}{2})^2 - 2 (z- \frac{x t}{\sqrt {2}} +
\frac {y t}{2}) -ty = 0;
\label{extrz}
\end{equation}
\begin{equation}
\frac {t^2 y}{4
(z- \frac{x t}{\sqrt {2}} + \frac {y t}{2})^2} = y;
\label{extry}
\end{equation}

\begin{equation}
x = \frac {t}{4
(z- \frac{x t}{\sqrt {2}} +  \frac {y t}{2})} + \frac{t^2 y} {8
(z- \frac{x t}{\sqrt {2}} +  \frac {y t}{2})^2}
\label{extrx}
\end{equation}

The solution of the equation ~(\ref{extrz}) gives
$$
z- \frac{x t}{\sqrt {2}} +  \frac {y t}{2} \to \frac {1}{4} $$
if $y\to 0$. So, the Eq.~(\ref{extry}) gives the critical
temperature $T_c$ defined by $4t_c^2 = 1$.

For $T>T_c$ the solution of the equations
~(\ref{extrz}) --~(\ref{extrx}) is
$$y = 0, x = t \sqrt {2}, z = \frac{1}{4} + t^2$$
and the replica symmetric free energy ~(\ref{frep}) coincides
with that given by ~(\ref{fav1}).
For $T<T_c$ we obtain
$$y = 2 t - 1, x = t \sqrt {2}, z = t$$
and the replica symmetric free energy ~(\ref{frep}) coincides
with that given by ~(\ref{fav2}).
Following ~\cite{KTJ} it is easy to show that the behavior of
the order parameters is identical in both approaches.
For standard glass order parameter $q$ we have $q = y/t = 2 -
1/t$.

So, the replica symmetric solution for the 3-state "spherical"
Potts glass coincides with the exact solution for this model
obtained by use of the known properties of large random
matrices so that no replica symmetry breaking is needed. This
case is in close analogy with the case of spherical version of
the SK model.

It is worth making some remarks. First,
it is interesting to emphasize that the exact solution
is the replica symmetric one, although there is no reflection
symmetry in the model.
The 3-state spherical Potts glass behaves like spherical
SK glass and not like the
 $p$-spin spherical model. So,
one can conclude that it is not the reflection symmetry
that define the replica  solution in spin glasses.

 Second, the model proposed in this paper permit to consider
 correctly the dynamics of glass transition.

Third, it is obvious that the
regular case can be solved exactly, too. Following ~\cite{BerK}
we can consider different lattices and short-ranged
interactions.  For the case of infinite-ranged regular
interaction ($J_{ij}=G$) analogous to that of Kac ~\cite{kac}
model we obtain directly from the partition sum ~(\ref{zz})
\begin{equation}
f(T) = -T -2 G -2 T \ln {2}, T>T_c,
\label{f1}
\end{equation}

\begin{equation}
f(T) = -4 G + T \ln {\frac{G}{2T}}, T<T_c,
\label{f2}
\end{equation}
with a discontinuity in specific heat at $T_c$ ($T_c=2G$).

This work was supported in part by the Russian Foundation for
Basic Research (Grants No. 02-02-16621 (NVG and EET) and No. 02-02-16622 (VNR))
and NWO-RFBR grant No. 04-01-89005.

\end{document}